\documentclass[prl,aps,showpacs,twocolumn,superscriptaddress,notitlepage]{revtex4}
\usepackage{amsmath}
\usepackage{graphicx}
\usepackage{bm}
\usepackage{amsbsy}
\usepackage{amsfonts}
\usepackage{amsthm}
\usepackage[dvips]{color}
\usepackage{epsfig}
\usepackage{graphicx}

\begin{document}

\title{Demonstrating macroscopic entanglement based on Kerr non-linearities requires extreme phase resolution}

\author{Roohollah Ghobadi}
\address{Institute for Quantum Information Science and
Department of Physics and Astronomy, University of Calgary,
Calgary T2N 1N4, Alberta, Canada}
\author{Sadegh Raeisi}
\address{Institute for Quantum Computing, University of Waterloo, Ontario N2L 3G1, Canada}
\author{Christoph Simon}
\address{Institute for Quantum Information Science and
Department of Physics and Astronomy, University of Calgary,
Calgary T2N 1N4, Alberta, Canada}

\begin{abstract}
Entangled coherent states, which can in principle be created using strong Kerr non-linearities, allow the violation of Bell inequalities for very coarse-grained measurements. This seems to contradict a recent conjecture that observing quantum effects in macroscopic systems generally requires very precise measurements. However, here we show that both the creation of the required states and the required measurements rely on being able to control the phase of the necessary Kerr-nonlinearity based unitary operations with extreme precision. This lends support to the idea that there is a general principle that makes macroscopic quantum effects difficult to observe, even in the absence of decoherence.
\end{abstract}
\maketitle

What does it take to observe macroscopic quantum effects? It is well known that one essential requirement is to minimize the impact of decoherence due to the coupling of the system to its environment \cite{zurek}. However, this is not necessarily sufficient. Already in 1979, Mermin \cite{mermin} showed that in order to obtain a violation of Bell's inequality for a singlet state of two large spins $J$, the angular resolution of the direction chosen for the spin measurement has to scale as $\frac{1}{J}$, which becomes more and more difficult for increasing $J$. Later, Peres showed that the resolution with which the spin projection values are measured also has to be high \cite{peres}. More recently it was shown \cite{entlaser} for closely related multi-photon singlet states that their entanglement can be demonstrated if photon counting measurements have a resolution better than $\sqrt{N}$, where $N$ is the total number of photons. Most recently, Ref. \cite{raeisi1} showed that for multi-photon states based on amplifying one half of an initial two-photon entangled state, micro-macro entanglement can be demonstrated only if photons can be counted with single-photon level precision. In Ref. \cite{raeisi1} it was conjectured that showing macroscopic quantum effects generally requires highly precise measurements, even in the absence of decoherence.

The multi-photon states considered in Refs. \cite{entlaser,raeisi1} can be created by $\chi_2$ non-linearities with a classical pump field, or more formally by generalized squeezing transformations. The associated Heisenberg equations of motion correspond to a linear mixing of creation and annihilation operators. This may lead one to question the generality of a conjecture based on such a relatively special class of states. The question is made more urgent by the results of Ref. \cite{jeongcoarse}, which showed that using a Kerr ($\chi_3$) non-linearity, for which the dynamics of field operators is also non-linear, it is possible to implement states and measurements that allow one to violate a Bell inequality using very coarse-grained homodyne detection. Does this mean that higher-order non-linearities make it fundamentally easier to observe macroscopic quantum effects? Similar questions concerning the usefulness of non-linearities have been raised in quantum metrology \cite{mitchell}.

Here we show that in the present context there is a significant price to pay. We are not referring to the practical difficulty of implementing strong Kerr non-linearities. While this is still an open challenge, there are several promising recent proposals \cite{rydberg,rispe}. In the spirit of the above discussion, we are also not concerned with the high sensitivity of the relevant multi-photon states to photon loss. Loss is due to the coupling of the system to its environment, whereas here we are interested in fundamental limits to the observability of macroscopic quantum effects even when the system is completely isolated. We show that there is a difficulty that is - in a sense - complementary to the precision requirements on photon number measurements discussed in Refs. \cite{entlaser,raeisi1}. Namely, the {\it phase} of unitary operations involving the Kerr non-linearity has to be extremely well defined. The required phase precision scales like $N^{-\frac{3}{2}}$, where $N$ is the mean number of photons.

\begin{figure}
\centering{}\includegraphics[width=0.8 \columnwidth]{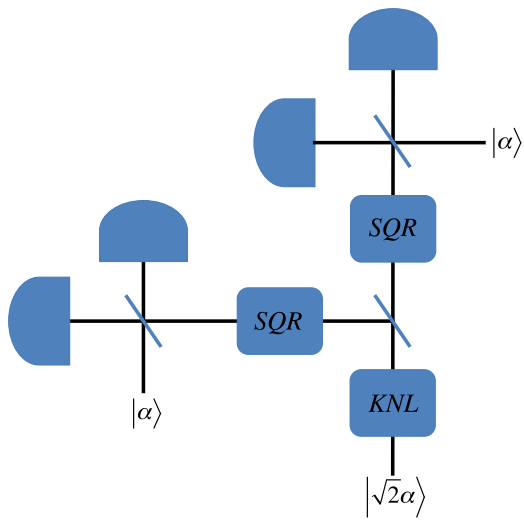}
\caption{Scheme to observe macroscopic entanglement with very coarse-grained measurements. The coherent state $|\sqrt{2} \alpha\rangle$ (with $\alpha \gg 1$) is transformed into a macroscopic superposition of coherent states Eq. (\ref{cat}) using a Kerr non-linear operation (KNL). The superposition is transformed into a maximally entangled state of coherent state qubits Eq. (\ref{ent}) using a beam splitter. General single-qubit rotations (SQR) can be implemented using the Kerr nonlinearity following Ref. \cite{jeongkim}. Detection in the coherent state qubit basis is done via interference with a coherent state $|\alpha\rangle$ on a beam splitter, followed by detectors that only need to distinguish the bright state $|\sqrt{2}\alpha\rangle$ from the vacuum. This setup would in principle allow the observation of Bell inequality violations (for example), thus demonstrating macroscopic entanglement.}
\label{fig1}
\end{figure}

We will begin by describing a conceptually simple scheme based on Kerr non-linearities that in principle allows the coarse-grained detection of macroscopic entanglement, see Fig. 1. Our scheme can be seen as a simplified version of the proposal of Ref. \cite{jeongcoarse}. It is different from that proposal both concerning the state that is used and the final measurements that are performed, but the phase precision requirements shown here apply to Ref. \cite{jeongcoarse} as well. We use the interaction Hamiltonian for a Kerr non-linearity,
\begin{equation}
H_{KNL}= (a^{\dagger} a)^2,
\end{equation}
where we have set the coupling constant equal to one for simplicity and $a$ is the annihilation operator for the relevant mode. We first use the interaction for a time $t=\frac{\pi}{2}$ to create a superposition of coherent states from an initial coherent state $|\sqrt{2} \alpha\rangle$ following Ref. \cite{yurke},
\begin{equation}
|\psi_\frac{\pi}{2}\rangle=e^{-i\frac{\pi}{2} H_{KNL}} |\sqrt{2}\alpha\rangle=\frac{e^{-i\pi/4}}{\sqrt{2}}(|\sqrt{2}\alpha\rangle+i|-\sqrt{2}\alpha\rangle),
\label{cat}
\end{equation}
where we are interested in the regime $\alpha \gg 1$ (we will take $\alpha$ to be real for simplicity). Sending this state onto a 50/50 beam splitter creates an entangled superposition of coherent states \cite{sanders},
\begin{equation}
\frac{e^{-i\pi/4}}{\sqrt{2}}(|\alpha\rangle_A |\alpha\rangle_B+i|-\alpha\rangle_A |-\alpha\rangle)_B,
\label{ent}
\end{equation}
where we have introduced two parties A and B corresponding to the two modes after the beam splitter. For $\alpha \gg 1$ this can be seen as a maximally entangled state of ``coherent-state qubits'' with basis states $|\alpha\rangle$ and $|-\alpha\rangle$ that are almost orthogonal. In order to measure an entanglement witness such as a Bell inequality on such a state we require two more ingredients, namely single-qubit rotations and measurements in the qubit basis. In Ref. \cite{jeongkim} it was shown that arbitrary single-qubit rotations can be implemented by combining the same Kerr non-linear operation that was used to create the initial coherent state superposition in Eq. (\ref{cat}) with small displacements in phase space, where the latter can be implemented using a strongly asymmetric beam splitter; in particular see Eqs. (9), (10) and (5) of Ref. \cite{jeongkim}.

The final missing ingredient is then the measurement in the qubit basis, i.e. a measurement that allows one to distinguish the states $|\alpha\rangle$ and $|-\alpha\rangle$. Such a measurement can be performed by interfering the state of light under consideration with an auxiliary coherent state $|\alpha\rangle$ on a 50/50 beam splitter \cite{branciard}. This corresponds to the transformations $|\alpha\rangle |\alpha\rangle \rightarrow |\sqrt{2} \alpha\rangle |0\rangle$ and $|-\alpha\rangle |\alpha\rangle \rightarrow |0\rangle |\sqrt{2} \alpha\rangle$ respectively. That is, for the input state $|\alpha\rangle$, the first output mode of the beam splitter will contain a strong coherent beam and the second output mode will be dark, whereas for the input state $|-\alpha\rangle$ the opposite is the case. The two states can then easily be distinguished by highly coarse-grained photon counting because all that is required is to distinguish a very bright state from the vacuum. The use of a strong Kerr non-linearity thus makes it possible to avoid the high-resolution requirement discussed in Ref. \cite{raeisi1}, at least as far as photon counting is concerned. This confirms the result of Ref. \cite{jeongcoarse} that it is {\it in principle} possible to observe a violation of Bell's inequality with very coarse-grained measurements.

\begin{figure}
\epsfig{file=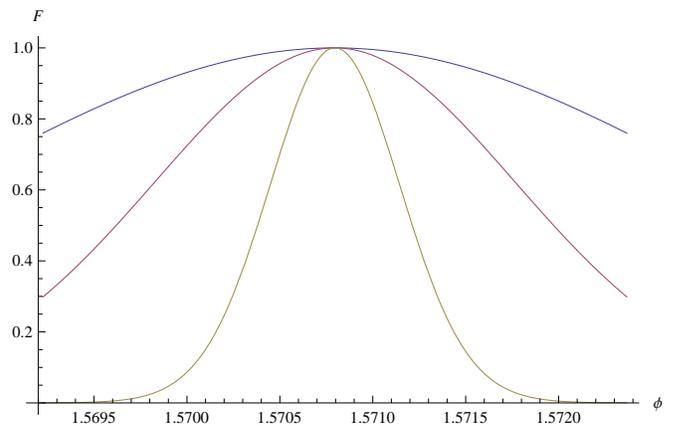,width=\columnwidth}
\caption{Fidelity of the state $|\psi_{\phi}\rangle$ of Eq. (\ref{psiphi}) relative to the ideal state $|\psi_{\frac{\pi}{2}}\rangle$ of Eq. (\ref{cat}) as a function of $\phi$, for values of $\phi$ close to $\frac{\pi}{2}$ and for different values of $\alpha$. The curves correspond to mean photon numbers $N=2 \alpha^2$ equal to 30, 50 and 100 from top to bottom. The fidelity decays faster and faster as the mean number of photons increases.}
\end{figure}

We now show that there is a significant difficulty with this approach.
The Kerr operation of Eq. (\ref{cat}), which also intervenes in the single-qubit rotations following Ref. \cite{jeongkim}, requires very high phase resolution. In Fig. 1 we plot the fidelity $F=|\langle \psi_{\phi} | \psi_{\frac{\pi}{2}} \rangle |^2$, where
\begin{equation}
|\psi_{\phi}\rangle=e^{-i\phi H_{KNL}} |\sqrt{2}\alpha\rangle,
\label{psiphi}
\end{equation}
as a function of $\phi$, for values of $\phi$ in the vicinity of the ideal value of $\frac{\pi}{2}$. The figure shows that for $\alpha \gg 1$ the fidelity of the created state with respect to the desired superposition of coherent states of Eq. (\ref{cat}) decreases very quickly as $\phi$ moves away from $\frac{\pi}{2}$. Moreover, this decay is faster and faster for increasing values of $\alpha$.

The scaling of the necessary phase resolution with the mean photon number $N=2 \alpha^2$ can be understood analytically. One has
\begin{equation}
\langle \psi_{\phi}|\psi_{\frac{\pi}{2}}\rangle=e^{-N} \sum_{n=0}^{\infty} \frac{N^n}{n!} e^{i(\phi-\frac{\pi}{2}) n^2}.
\end{equation}
Using the Stirling expansion for $\ln n!$, defining $x=n-N$ and $\tilde{\phi}=\phi-\frac{\pi}{2}$, approximating the sum over $x$ by an integral, and keeping only the terms that are dominant in the limit of large $N$, one finds that this is proportional to
\begin{equation}
\int_{-\infty}^{\infty} e^{-\frac{x^2}{2N}} e^{i2N\tilde{\phi}x},
\end{equation}
where the proportionality factor can be inferred from the fact that the overlap is equal to one for $\tilde{\phi}=0$. Performing the integral gives a Gaussian distribution for $\tilde{\phi}$ whose width scales like $N^{-\frac{3}{2}}$, in good correspondence with the results shown in Fig. 1 \cite{zurekcats}. The need for extreme phase resolution (for $\alpha \gg 1$) applies both to our scheme and to the scheme of Ref. \cite{jeongcoarse}, which uses the Kerr nonlinearity in a similar way.

These results show that while it is in principle possible to observe macroscopic quantum effects with very coarse-grained photon {\it number} measurements in this system, one has to pay the price of increasingly precise (as the size of the system increases) {\it phase} control for the operations involving the Kerr non-linearity. This supports the idea that there may be a general principle that makes it hard to observe macroscopic quantum phenomena, even in the absence of environmentally induced decoherence. The precise form of this principle remains to be discovered. Comparing the present results to those of Ref. \cite{raeisi1} one is tempted to conjecture the existence of a number-phase trade-off (similar to an uncertainty relation), which would imply that observing quantum effects in macroscopic systems requires either very precise photon number measurements or very precise phase control. However, it should be noted that the phase considered here is that of a non-linear operation, which is a different concept from the phase observable that is complementary to photon number in several respects, including the fact that it is a control parameter and not an observable. Note that Mermin's result in Ref. \cite{mermin} concerned a control parameter, while the results of Refs. \cite{peres,entlaser,raeisi1} concerned the precision of measurements. It may be possible to gain more insight into these questions by studying further examples. In particular, it would be interesting to find cases where there are trade-offs between the requirements for number and phase precision, and also between control precision and measurement precision.

\begin{acknowledgments}
We thank C. Branciard, N. Gisin, H. Jeong, T. Ralph, B. Sanders, and V. Scarani for useful discussions. This work was supported by AITF and NSERC.
\end{acknowledgments}



\end{document}